\documentclass[12pt,a4paper]{article}
\usepackage{graphicx,amsmath,amsfonts}
\def \ba{\begin{eqnarray}}\def\ea{\end{eqnarray}}
\def\bc{\begin{center}}\def\ec{\end{center}}
\def\nn{\nonumber\\}
\title{\Large \bf The electromagnetic effects in  $K_{e4}$ decay.}
\author{\large \bf S.R.Gevorkyan\footnote{On leave of absence from
Yerevan Physics Institute},~~A.N.Sissakian,~~A.V.Tarasov,\\
\bf ~~H.T.Torosyan,~~O.O.Voskresenskaya\footnote{On leave of absence
from Siberian Physical Technical Institute}}

\begin{document}
\maketitle \bc Joint Institute for Nuclear Research, 141980 Dubna,
Russia \ec

\begin{abstract}
 The final state interaction of pions in $K_{e4}$  decay allows  to
obtain  the value of the isospin  and angular momentum  zero  $\pi\pi$
scattering length $a_0^0$. We take into account the  electromagnetic interaction
of pions and isospin symmetry breaking effects caused  by  different
masses of neutral and charged pions and estimate  the  impact of these effects
on the procedure of scattering length  extraction from   $K_{e4}$ decays.
\end{abstract}

\section{Introduction}
 For many years  the decay
  \ba
   K^\pm \to \pi^+\pi^- e^\pm \nu
   \ea
 was considered as the cleanest method to determine the isospin and
angular momentum  zero scattering length $a_0^0$ ~\cite{ACGL}. At present the value of $a_0^0$ is
predicted by Chiral Perturbation Theory (ChPT)  with high precision~\cite{CGL}
  and its measurement with relevant accuracy  can provide  useful
constraints on the  ChPT Lagrangian. The appearance of
new precise experimental data ~\cite{pis,mas,bb,bat1} requires  approaches
which can take  into account the  effects  neglected up to now in
 extracting the scattering length from experimental data on $K_{e4}$ decays.\\
 The common  way  to  get the scattering length $a_0^0$
from the decay  probability   is based on the classical works
~\cite{CM,PT}.The transition amplitude for decay (1) can be written as
the product of the lepton and hadronic currents:
\ba
A=\frac{G_F\sin{\theta_c}}{\sqrt{2}}\langle \pi^+\pi^-|J^{\mu}_{had}|K^+\rangle
\langle e^+\nu_e|J_{\mu}^{lep}|0\rangle.
\ea
The leptonic  part of this  matrix element  is known
 exactly,while the hadronic part can be described by four hadronic
form factors\footnote{The form factor R is proportional to the
electron mass and thus it can not be extracted from $K_{e4}$ decay}  F,G,R,H ~\cite{PT}.
Making the partial-wave expansion of the hadronic
current with respect to the angular momentum of the dipion system
 the hadronic  form factors can be written in the following form:
\ba
F&=&f_se^{i\delta_s(s)}+f_pe^{i\delta_p(s)}
\cos{\theta_{\pi}}\nn
G&=&g_pe^{i\delta_p(s)};~~~H=h_pe^{i\delta_p(s)}
\ea
Here $s=M_{\pi\pi}^2$ is the square of  dipion invariant mass;
$\theta_{\pi}$ is the polar angle of the pion in the dipion rest frame measured
with respect to the flight direction of dipion in the K  meson rest
frame. The coefficients $f_s,f_p,g_p,h_p$ can be parameterized as
functions of pion momenta q in the dipion rest system  and of the invariant
mass of lepton pair $s_{e\nu}$ in the known way  \cite{AB}.
It is widely accepted that the  s and p- wave phases $\delta_s,\delta_p$    coincide  with
the corresponding phases  in elastic $\pi\pi$ scattering  (Fermi---Watson theorem ~ \cite{W})
 and can be  related to the scattering lengths using the set of Roy equations~\cite{ACGL}. \\
Nevertheless the different masses of charged and neutral pions  lead to the  isospin symmetry breaking
~\cite{G1,GSTTV1,G2}  and require the new approach to connect   the phases with
scattering lengths.\\
Another isospin symmetry  breaking effect   is the  electromagnetic interaction in the dipion system
~\cite{G2, G3, GSTTV2}, which would has impact on the value of scattering length  extracted from $K_{e4}$ decay rates.
In the present work  we develop  the approach allows one to
take into account the electromagnetic interaction in the dipion system and estimates its impact on
the value of scattering lengths extracted from  $K_{e4}$ decay.
\section{Isospin symmetry  breaking due to pions mass difference}
The s-wave phase shift $\delta_s$   has an impact only on axial  form factor F,
whereas the axial form factors G and vector form factor
H depend only on p-wave phase shift $\delta_p$. If one confins  by  s and p waves,
the inelastic process $\pi^0\pi^0\to \pi^+\pi^-$  and the reversed   one are  forbidden due to identity
of neutral pions in  l=1  state. Thus  inelastic transitions can change only the first term
in the form factor F, relevant to production of  pions in s-wave.\\
In one loop approximation of nonperturbative effective field theory
(see e.g. \cite{G4}) the decay amplitude relevant  to dipion  in the state with I=l=0 reads:
\ba
T=T_1(1+ik_ca_c(s))+ik_na_x(s) T_2.
\ea
Here  $T_1,T_2$ are the so called ``unperturbed'' amplitudes ~\cite{C,CI}  corresponding
to the decays with  charged and neutral dipions in the final state.\\
$k_n=\frac{\sqrt{s-4m_0^2}}{2},~~k_c=\frac{\sqrt{s-4m_c^2}}{2}$
are the pion  momenta in  the $\pi^0\pi^0$ and $\pi^+\pi^-$ systems  with the
same invariant mass s= $ M_{\pi\pi}^2$. The real functions   $a_c(s),a_x(s)$ are
 relevant to   elastic scattering $\pi^+\pi^-\to\pi^+\pi^-$
and charge exchange reaction $\pi^0\pi^0\to\pi^+\pi^-$ .\\
 In the case of isospin symmetry they  can be expressed through the  s-wave "amplitudes"  with certain
 isospin  $a_0(s),a_2(s)$,which at  threshold  are equal to relevant scattering lengths $a_0^0,a_0^2$.
 In the case of isospin  symmetry breaking  we adopt the relations followed from  ChPT~\cite{G4}:
\ba
a_c(s)&=&\frac{2a_0(s)+a_2(s)}{3}(1+\eta);\nn
a_x(s)&=&\frac{\sqrt{2}}{3}(a_0(s)-a_2)(1+\frac{\eta}{3});
\eta=\frac{m_c^2-m_0^2}{m_c^2}
\ea
In the isospin symmetry limit ($k_c=k_n=k; \eta=0$)  a simple relation takes place between the
``unperturbed'' amplitudes $ T_1=\sqrt{2}T_2 $ , which   follows from the rule $\Delta I=1/2$
for semi-leptonic decays.  In this limit it is easy to obtain:
\ba
T=T_1(1+ika_0(s))=T_1\sqrt{1+k^2a_0(s)^2}e^{i\delta_0^0}.
\ea
This equation is nothing else than the Fermi-Watson theorem~\cite{W}  for
the $\pi\pi$  interaction in the final states.\\
In the general case using the expressions  (4),(5)  and  relations between the s-wave "amplitudes"
 and relevant phases :
\ba
\tan\delta_s(s)=k_ca_c(s);  \tan\delta_0^0=k_c  a_0(s);\tan\delta_0^2=k_c  a_2(s)
\ea
after a bit algebra it is easy to obtain:
\ba
\delta_s&=&\arctan(A_s\tan\delta_0^0+B_s\tan\delta_0^2)\nn
A_s&=&\frac{2(1+\eta)+\lambda(1+\frac{\eta}{3})}{3};
B_s=\frac{(1+\eta)-\lambda(1+\frac{\eta}{3})}{3};
\lambda=\frac{k_n}{k_c}
\ea
Another isospin breaking effect which can be important in  the procedure of the scattering lengths extraction
from the experimental data on $K_{e4}$ decay,  is the Coulomb interaction between the
charged pions~\cite{G2,G3,GSTTV2} The widely spread wisdom is that in order to take this  effect  into
account it is  sufficient to multiply the square of matrix element   (2) by Gamov factor
\ba
G=\frac{2\pi \xi}{1-e^{-2\pi \xi}}; \xi=\frac{\alpha}{\upsilon};\upsilon=\frac{\sqrt{1-4\beta}}{1-2\beta};\beta=\frac{2k_c}{\sqrt{s}}
\ea
Here $\upsilon$ is  the relative  velocity in the dipion system and
    $\alpha=\frac{e^2}{4\pi}$ is the fine structure constant.\\
Later on we  show that  besides this multiplier the electromagnetic
interaction between pions  also change the expression (8) for the strong
phase and add the proper  Coulomb phase.
\section{Electromagnetic interaction in $\pi\pi$ system}
In order to take into account the  electromagnetic interactions between pions, we take an advantage
of the trick  successfully   used in ~\cite{GTV}.To switch on the electromagnetic interaction, we
 replace the charged pion momenta $k_c$ in (7)
by a logarithmic derivative of the pion wave function in the Coulomb potential at the
boundary of the strong field  $r_0$ :
 \ba
 ik_c\to\tau=\frac{d\log[G_0(k r)+iF_0(k r)]}{dr}\biggl |_{r=r_0}.
 \ea
Here $F_0,G_0$ are the regular and irregular solutions of the Coulomb problem.\\
In the region $ kr_0\ll 1$, where the electromagnetic effects are
significant,  this expression can be simplified:
\ba
\tau &=& ik-\alpha
m\left[\log(-2ikr_0)+2\gamma+\psi(1-i\xi)\right]=Re~\tau+i~Im~\tau\nn
Re~\tau &=&-\alpha m\left[\log(2kr_0)
+2\gamma+Re\,\psi(1-i\xi)\right];\nn
 Im~\tau&=&\frac{\pi k\xi}{\sinh{\pi\xi}}e^{\pi\xi}
\ea
Here   $\gamma=0.5772$ is Euler constant
and   $\psi(z)=\frac{d\log\Gamma(z)}{dz}$ digamma  function. \\
Using the above  relations one can express the modified
phase for  $\pi^+\pi^-$ state  (I=l=0)    through the known ~\cite{ACGL}
phases   $\delta_0^0,\delta_0^2$ .\\
Representing  the modified s-wave  phase   as a sum of strong  $\delta_{str}$ and electromagnetic
 $\delta_{em}$   terms,  we obtain:
\ba
\tilde\delta_s&=&\delta_{str}+\delta_{em}\nn
\delta_{str}&=&\arctan(A_{em}\tan{\delta_0^0}+B_{em}\tan{\delta_0^2});
\delta_{em}=\arctan(\frac{\alpha}{\upsilon});\nn
A_{em}&=&\frac{2G (1+\eta)+\lambda(1+\frac{\eta}{3})}{3};~~~
B_{em}=\frac{G (1+\eta)-\lambda(1+\frac{\eta}{3})}{3}
\ea
Let us note that, whereas the electromagnetic phase $\delta_{em}$ has a common textbook
form ~\cite{LL},  the strong phase is essentially modified by electromagnetic effects
(the Gamov factor G  in $\delta_{str}$) as well as by   isospin symmetry breaking effects
provided by pions mass difference.\\
Using  the same approach one can show that the modified  p-wave phase
reads:
\ba
\tilde{\delta_p}=\arctan \left(G(1+\frac{\alpha^2}{\beta^2}) \tan{\delta_1^1}\right).
\ea

 Setting   $a_0^0=0.225{m_c}^{-1};a_0^2=-0.03706{m_c}^{-1}$  and using the relevant   phases
 $\delta_0^0,\delta_1^1$ from Appendix D  of ~\cite{ACGL}, we   calculated  the modified
 phases differences $\delta=\tilde\delta_s-\tilde\delta_p$
as a function of the invariant mass of dipion $M_{\pi\pi}$.\\
The dashed line on Fig.1 corresponds to exact isospin symmetry limit  $m_0=m_c;\alpha=0.$
The solid line gives  the  dependence of modified phases difference accounting for   all isospin breaking effects.
  The  experimental data  are from~\cite{mas}.\\
 The  considered above isospin breaking effects change remarkably   $\delta$  and would
 have impact on the values of scattering lengths extracted from experimental data.\\
In table 1 we cite  $\delta$ as a function of dipion invariant mass $M_{\pi\pi}$
in respect to different isospin breaking corrections. This allows one to estimate separately
the contribution of considered above effects.

\begin{table}[h]
\vspace{1cm}
\caption{The impact of  considered corrections on phase difference $\delta=\delta_s-\delta_p$:
1) Standard  case~\cite{ACGL}  with $a_0^0=0.225{m_c}^{-1};  a_0^2=-0.03706{m_c}^{-1}$
 2) With charge exchange process  $\lambda=\frac{k_n}{k_c}$
 3) With parameter  $\eta$ (expression (5))
4) With electromagnetic interaction.  5) With the additional Coulomb phase }
\begin{center}
\begin{tabular}{|c|c|c|c|c|c|}
\hline\hline
$M_{\pi\pi}$ & 1 & 2 & 3 & 4 & 5 \\
\hline
\textbf{0.285} &\textbf{0.048} &\textbf{0.059}&\textbf{0.061} &\textbf{0.063} &\textbf{0.082}\\
\hline
\textbf{0.300}&\textbf{0.096}&\textbf{0.103}&\textbf{0.108} &\textbf{0.110} &\textbf{0.122}\\
\hline
\textbf{0.315} &\textbf{0.134} &\textbf{0.140}&\textbf{0.147} &\textbf{0.149} &\textbf{0.159}\\
\hline
\textbf{0.330}&\textbf{0.170}&\textbf{0.175}&\textbf{0.184} &\textbf{0.186} &\textbf{0.195}\\
\hline
\textbf{0.345} &\textbf{0.205} &\textbf{0.210}&\textbf{0.220} &\textbf{0.223} &\textbf{0.231}\\
\hline
\textbf{0.360} &\textbf{0.239} &\textbf{0.244}&\textbf{0.256} &\textbf{0.259} &\textbf{0.267}\\
\hline
\textbf{0.375}&\textbf{0.274}&\textbf{0279}&\textbf{0.292} &\textbf{0.296} &\textbf{0.304}\\
\hline
\textbf{0.390} &\textbf{0.309} &\textbf{0.314}&\textbf{0.328} &\textbf{0.333} &\textbf{0.340}\\
\hline
\end{tabular}
\end{center}
\end{table}
\newpage
\section{Conclusions}
The  isospin symmetry breaking corrections considered above  increase
the phase difference $\delta$. Their contribution is the largest near
the threshold, but they are essential even far from it.\\
The $K_{e4}$ decay amplitude in  the real world with  isospin symmetry breaking
  depends  on two scattering  lengths  $a_0^0,a_0^2$,   unlike the common approach.
 The  proposed approach allows one to  extract the  values of scattering lengths
with  higher accuracy than in standard approximation. \\
The authors thank   V.D. Kekelidze and D.T. Madigozhin for
useful discussions and support. We are grateful to J.Gasser and A.Rusetsky  for useful
comments and friendly criticism, which assist   to  improve the present
work and our understanding of considered above problems.

\newpage
\begin{figure}[ht]
\begin{center}
\includegraphics[scale=1]{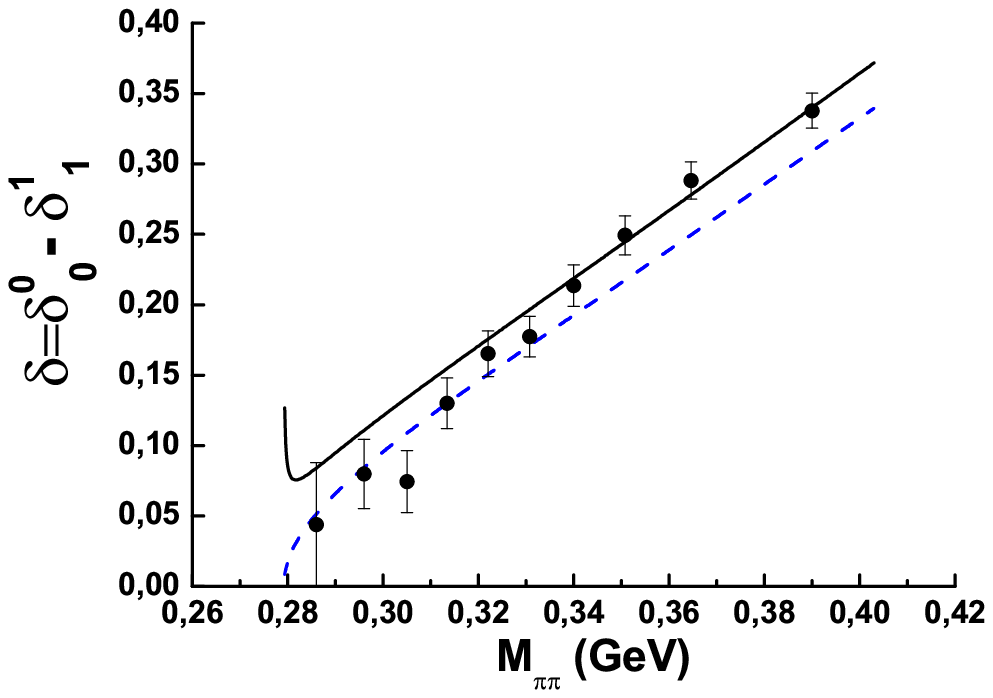}
\caption{\Large The dependence of  phases difference $\delta=\delta_s-\delta_p$  on  dipion invariant mass
in the exact isospin symmetry case (dashed line)  and with  all  isospin symmetry breaking corrections taken into account
(solid line).  }
\end{center}
\end{figure}

\end{document}